# Looking Back: The Changing Landscape of Abortion Care in Louisiana

Mayra Pineda-Torres, PhD, and Yana van der Meulen Rodgers, PhD

**Abstract:** This article examines how COVID-19 and the *Dobbs* decision have impacted abortion services in Louisiana. COVID-19's introduction into an already restrictive landscape of abortion policies intensified the barriers that providers and communities faced, with disproportionate impacts on Black and Hispanic abortion seekers. The 2022 *Dobbs* decision marked the immediate enactment of Louisiana's abortion ban, resulting in even greater difficulties in accessing abortion services. Concerns raised by Roberts et al. (2021) about the negative effects of clinic closures have only grown since their prescient study.

**I. Introduction**

Access to abortion, a healthcare service already long under siege from opponents, became even more tenuous as a result of widespread shutdowns and shelter-in-place orders during the COVID-19 pandemic. Service providers also experienced adverse effects, including shortages of personal protective equipment to ensure the safety of essential workers. COVID-19's introduction into an already restrictive landscape of abortion policies intensified the barriers that providers and communities faced, disproportionately impacting Black and Hispanic abortion seekers (Wolfe and Rodgers 2021).

In August 2021, Roberts et al. published a research article in the *American Journal of Public Health* showing that pandemic-related lockdowns and legal restrictions led to a 31 percent decline in the number of abortions per month in Louisiana but an increase in the likelihood of obtaining a second-trimester abortion. In this editorial, we examine how COVID-19 and the *Dobbs* decision have impacted abortion services in Louisiana since the publication of this influential article. Concerns raised by Roberts et al. about the negative effects of clinic closures have only grown since their prescient study.

**II. Covid-Related Impacts on Abortion Services in Louisiana**

COVID-19 exacerbated the barriers people face in trying to access or practice abortion care. Across the country, social distancing requirements and the lack of childcare in the midst of school shutdowns placed limits on staffing capacity and the number of patients that clinics could schedule. Already experiencing a hostile legal environment before the pandemic, clinics in Louisiana faced additional threats as the state joined twelve others in designating abortion services as "nonessential" early in the pandemic (Ruggiero et al. 2020). Officials argued that restricting abortions would free up medical supplies and personnel by postponing elective procedures until the end of the crisis. The Covid abortion bans were designed to reduce the number of abortions or force people out of state for abortion services, further adding to the monetary and time costs of obtaining an abortion.

Abortion is a time-sensitive service both in terms of health and in terms of legal restrictions. Forcing those who are pregnant to delay an abortion may endanger their physical health if the individual has a high-risk pregnancy or has complications from the procedure. Even though abortion is very safe, complications such as infection, hemorrhage, or uterine perforation are more common in second-trimester than first-trimester abortions (Grossman et al. 2008). Later or more complicated abortions are also more expensive and may entail more adverse mental health effects (Coast et al. 2021). Delays could also extend the pregnancy to the point of fetal viability (designated as 20 weeks according to Louisiana law), after which most states prohibit abortions except to protect the life and health of the individual. The inclusion of abortion on the list of nonessential services was legally contested, with litigation in Louisiana and most other states resulting in abortion services remaining accessible.

Roberts et al. (2021) show that the monthly number of abortions requested by Louisiana residents in Louisiana clinics declined by 31 percent during the pandemic onset, with no evidence of an increase in out-of-state abortions that could compensate for this decline. The timing of abortions also changed since the likelihood of a second-trimester abortion increased. During the pandemic onset, Louisiana had only three open clinics, with only one or two of those having available appointments in any given month. This lack of service availability reflected a median wait time of more than two weeks, higher than in neighboring states. Louisiana also experienced a decrease in the proportion of abortions that are medication abortions (as opposed to aspiration abortion services), which the authors attribute to service availability in a facility that provided more aspirations than medication abortions. The relative decline may also be explained by the fact that Louisianans were forced to delay procedures (as demonstrated by the increase in second-trimester abortions), likely pushing them past the point at which medication abortion was an option.

Subsequent evidence in Berglas et al. (2022) indicates that Louisiana's decline in abortions, delay in abortion timing, and change in method were mainly concentrated among residents in areas where abortion care was disrupted. Moreover, subsequent data released by the Guttmacher Institute indicate that the percentage of Louisiana residents obtaining abortions who traveled out of state increased from 13% in 2019 to 21% in 2020 (Guttmacher Institute 2023). These studies show that Louisiana's pandemic abortion ban meaningfully disrupted people's ability to obtain abortions. As a result of the *Dobbs* decision, disruptions in abortion care have only worsened.

**III. The Dobbs Decision and Repercussions for Louisiana**

On June 24, 2022, the U.S. Supreme Court overturned *Roe v. Wade* and most aspects of *Planned Parenthood v. Casey* in the case *Dobbs v. Jackson Women's Health Organization*, which has left the legality of abortion up to the states. Before the *Dobbs* decision, most states had already implemented a series of state-level legal restrictions affecting both abortion seekers and providers, including parental consent for minors, targeted restrictions on abortion providers, mandatory pre-abortion counseling, pre-abortion waiting periods and testing requirements, physician-only laws, restrictions on medication abortion and telehealth, and insurance bans.

These restrictions resulted in clinic closures, fewer available appointments, and longer travel times and distances to obtain an abortion (Fischer et al. 2018; Lindo et al. 2020; Venator and Fletcher 2021). They also increased the monetary costs of abortion, which is already an expensive procedure and relatively difficult to finance for individuals with low incomes (Lindo and Pineda-Torres 2021). Federal and state restrictions on public funding for abortion costs, including the Hyde Amendment, further raised the out-of-pocket costs of abortions and increased the difficulty in accessing services for individuals with low incomes. Restrictions on private insurance plans that included abortion services had similar outcomes for abortion accessibility and costs. Abortion restrictions also adversely impacted child health, as shown in Foster et al.'s (2018) comparison of individuals who received abortions with individuals who were denied abortions due to state regulations: people who were able to delay childbirth until they had greater economic and emotional security were able to raise their children in relatively better economic circumstances, with fewer indicators of delayed child development.

The *Dobbs* decision allowed states not just to restrict abortion but also to ban abortion outright, even before viability. As of January 2024, abortion has been completely banned in 14 states, including Louisiana, and banned at an early gestational age in another two states. Note that several organizations have abortion law trackers, including the New York Times, Center for Reproductive Rights (CRR), Guttmacher Institute, and Kaiser Family Foundation. According to the Guttmacher Institute, Alabama, Arkansas, Idaho, Indiana, Kentucky, Louisiana, Mississippi, Missouri, North Dakota, Oklahoma, South Dakota, Tennessee, Texas, and West Virginia have all banned abortion, and Georgia and South Carolina have implemented six-week abortion bans.

Many people must travel long distances to access abortion services. Nationally, the mean travel distance to access an abortion provider was 27.8 minutes pre-*Dobbs* and 100.4 minutes

post-*Dobbs*, with most of the additional time incurred by people living in and surrounded by ban states in the South and Midwest, especially Louisiana and Texas (Rader et al. 2022). Before the *Dobbs* decision, 15 percent of reproductive-age women lived over one hour away from an abortion provider; after *Dobbs,* that share rose to 33 percent (Rader et al. 2022).

In Louisiana, the average distance to the nearest facility was 47 miles as of May 1, 2022 (Myers, 2023). However, a year later in the wake of the *Dobbs* decision, the average distance for Louisiana residents increased to 456 miles—a ten times increase in distance in one year (Figure 1). The state government further restricted access to abortion in June 2022 by making it a crime to receive mifepristone, the first of two drugs in a medication abortion, by mail, even though the FDA had lifted restrictions on mailing mifepristone. As a result of the state's restrictive laws, Myers (2023) predicts a decrease in Louisiana's abortion rate of 29 percent (unfortunately, data on post-Dobbs actual abortion rates for Louisiana are not available). Louisiana residents have never experienced such a detrimental decline in access to abortion facilities as the one they are currently experiencing after *Dobbs*. Even during the COVID pandemic, the average distance to the nearest abortion provider remained around 47 miles until the implementation of the state's abortion ban. In principle, people in Louisiana can obtain abortion services in other states. However, given the restrictive abortion landscape in neighboring states, distance acts as a barrier to abortion, disproportionately affecting individuals with low incomes who lack the time and financial resources to travel out of state. This detrimental access to abortion is calculated to have "trapped" 23 percent of Louisianian abortion seekers and to have increased births by 3.2 percent in Louisiana during the first six months of 2023 (Dench et al. 2023).

**IV. Moving Forward**

Distance does not provide the complete picture of people's ability to obtain abortion services. After the *Dobbs* decision, every state showed a higher request rate of medication abortion from a telemedicine service involving a licensed provider, with the largest increase among abortion ban states (Aiken et al. 2022). Louisiana experienced the highest increase in weekly requests, from 5.6 per 100,000 female residents (in September 2021-May 2022) to 14.9 (in June-August 2022), even though Louisiana prohibits the use of telemedicine to provide medication abortions. Therefore, the inability to access abortion services has obliged abortion seekers to rely on self-managed abortion with pills obtained online outside of the formal healthcare system. Although self-managed abortion with pills can be medically safe and effective, it may not be an option for all Louisiana abortion seekers, given financial constraints and the legal risks of this alternative. Moreover, medication abortion, whether it be through a telemedicine provider or self-managed, is most effective early in pregnancy and may not be safe or effective for many later abortion-seekers.

A major political shift on abortion, at either the state or the federal level, is necessary but may take many years. In the meantime, relaxing existing restrictions on medication abortions and increasing advocacy efforts around eliminating legislative roadblocks to telemedicine delivery methods will go a long way to getting people the treatment they need. However, while expanding access to medication abortion is extremely important, some people will still require or prefer in-clinic care. The risk of avoiding a comprehensive shift in abortion law is a greater likelihood of individuals being unable to access facility-based care or to self-manage their abortions with pills; instead, they may use other methods that are ineffective and potentially unsafe (Verma and Grossman, 2023). Fewer restrictions on medication abortions will also support abortion providers doing this work by reducing the various risks they face, including exposure to the COVID-19 virus and harassment at in-person clinics. Such measures have the potential to greatly reduce traditional racial, economic, and geographic barriers to abortion care.

**About the Authors**

Mayra Pineda-Torres is with the School of Economics, Georgia Institute of Technology, Atlanta, GA.

Yana van der Meulen Rodgers is with the Labor Studies and Employment Relations Department, Rutgers University, New Brunswick, NJ.

**Correspondence**

Correspondence to Yana Rodgers, 94 Rockafeller Road, Piscataway, NJ, 08854. Tel 848-932-4614, email yana.rodgers@rutgers.edu.

Acceptance Date: January 22, 2024

**Contributors**

MPT and YVR contributed equally to the conceptualization, writing, and editing of the article.

**Acknowledgments**

The authors thank Mary Northridge, Alfredo Morabia, and Daniel Fox for the opportunity to contribute to the AJPH's Looking Back initiative, and they thank Elizabeth Sully, Jocelyn Finlay, and three anonymous reviewers for their helpful suggestions.

**Conflicts of Interest**

The authors have no potential or actual Conflicts of Interest to declare.

Venator J, Fletcher J. Undue burden beyond Texas: An analysis of abortion clinic closures, births, and abortions in Wisconsin. *Journal of Policy Analysis and Management*. 2021 Jun;40(3):774-813.

Verma N, Grossman D. Self-Managed Abortion in the United States. Current Obstetrics and Gynecology Reports. 2023 Mar 7:1-6.

Wolfe T, Rodgers Y. Abortion during the COVID-19 pandemic: racial disparities and barriers to care in the USA. *Sexuality Research and Social Policy*. 2021:1-8.
10

**Fig. 1 Average distance from a county in Louisiana to the nearest abortion facility**

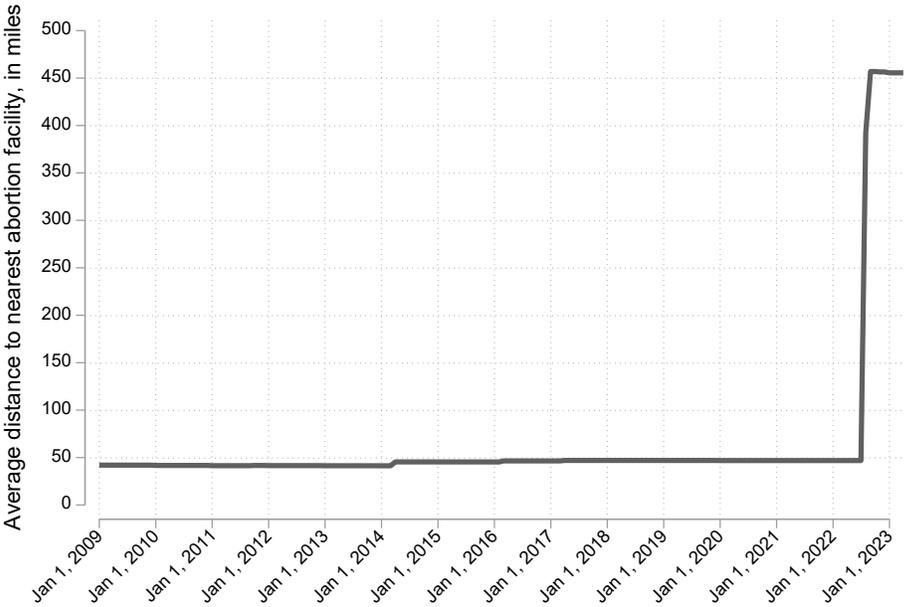

Note: Own elaboration using information on county-level distance to the nearest abortion facility from Myers (2021). The averages are weighted by the population of women aged 15-44 in the origin county.

11